\documentclass[conference]{IEEEtran}
\IEEEoverridecommandlockouts

\usepackage{amsmath,amssymb,amsfonts}
\usepackage[numbers]{natbib}
\usepackage{algorithmic}
\usepackage{graphicx}
\usepackage{textcomp}
\usepackage{multirow}
\usepackage[linesnumbered, ruled]{algorithm2e}
\usepackage{paralist}
\usepackage{booktabs}
\usepackage{url}
\usepackage{xcolor}
\usepackage{colortbl}
\usepackage{fancyvrb}
\usepackage{subcaption}
\usepackage{refcount}
\usepackage[skins]{tcolorbox}

\definecolor{gr}{gray}{0.8}
\definecolor{lightgray}{rgb}{0.83, 0.83, 0.83}

\usepackage[binary-units]{siunitx}
\DeclareSIUnit{\pp}{\textup{pp}}
\SetAlFnt{\small}
\SetAlCapFnt{\small}
\SetAlCapNameFnt{\small}

\usepackage{xspace}
\newcommand{\app}{LogPurifier\xspace}

\usepackage[colorlinks, allcolors=blue]{hyperref}

\begin{document}

\title{Cleaning Logs for Downstream Tasks \\(Registered Report)}

\author{\IEEEauthorblockN{Zahra G. Yazdi\IEEEauthorrefmark{1}, Van-Hoang Le\IEEEauthorrefmark{1}, Nyyti Saarimäki\IEEEauthorrefmark{1}, Donghwan Shin\IEEEauthorrefmark{2}, Domenico Bianculli\IEEEauthorrefmark{1}  and Lionel Briand\IEEEauthorrefmark{3}}
	\IEEEauthorblockA{\IEEEauthorrefmark{1}University of Luxembourg, 
		Luxembourg, 
		Email: first.last@uni.lu}
	\IEEEauthorblockA{\IEEEauthorrefmark{2}University of Sheffield, 
		United Kingdom, 
		Email: d.shin@sheffield.ac.uk}
		
		\IEEEauthorblockA{\IEEEauthorrefmark{3}University of Ottawa, 
			Canada and the Research Ireland Lero  Centre for Software Research, University of Limerick, \\
			Email: lbriand@uottawa.ca
			}
		}

\maketitle

\begin{abstract}
\textit{Background}: Software systems generate logs during execution to record critical events and runtime information for troubleshooting and monitoring. However, in practice, logs often contain significant amounts of redundant and irrelevant information, which can negatively impact the performance of downstream analysis tasks, such as model inference and anomaly detection. 
\textit{Objective}: The objective of this study is to clean log data by identifying and removing free-standing messages—messages that are not relevant to the execution behaviors of interest and are interleaved with messages capturing the system’s functional behavior.
\textit{Method}: To address this objective, we propose \app, a task-agnostic log-cleaning approach based on dependency relationships between log message templates. The paper presents a plan for an empirical evaluation using a controlled experimental design to assess the impact of \app on the effectiveness and efficiency of two downstream tasks: model inference and anomaly detection.

\end{abstract}


\section{Introduction}
\label{sec:introduction}

Software logging is a well-established and widely adopted practice in software engineering that has been extensively studied in the literature~\cite{Batoun2024,9756253}. 
Log files generated during software execution provide valuable runtime information and support a wide range of downstream tasks, including model inference~\cite{Shin2022}, anomaly detection~\cite{Zhang2020}, testing~\cite{10664280}, and performance monitoring~\cite{Yao2020Log4Perf}. However, in practice, the usefulness of the logs is often limited by quality issues. Since logs are produced by developers, their quality varies with developer expertise, organizational logging guidelines, and expectations about future use of the logged information. Consequently, real-world logs are often inconsistent, large, and noisy~\cite{Zhang2020, Fu2014}.

Developers commonly rely on tools to analyze log data~\cite{Zhang2020}. However, both noise and large volumes of data create challenges for such tools: large volumes increase computational costs, while noise can confuse analysis models and reduce overall performance. Similar challenges are widely recognized in machine learning and statistical analysis, where data cleaning is considered an essential preprocessing step~\cite{ilyas2019data}.  

Cleaning log data, however, is not straightforward. Logs may contain several types of noise, such as overly detailed messages, repeated templates, or messages that are not related to the functional behavior of the system. These challenges are particularly problematic in black-box settings where source code is unavailable and distinguishing which log templates are relevant to downstream tasks becomes even more challenging.

This paper focuses on a specific form of log noise: \emph{free-standing} log messages. We define free-standing messages as log messages that do not depend on their predecessors or successors. In contrast, we refer to log messages that convey the system's functional behavior as \emph{regular messages}. Such messages are common in practice, as system logs routinely include operational events alongside execution-related logs, often interleaved within the same traces~\cite{8030620,Oliner2007}. A common example of a free-standing message is a periodic heartbeat message that tracks system memory usage as it reflects operational status rather than the execution workflow. When interleaved with regular messages, they can distort event sequences and degrade downstream task performance~\cite{8030620,Jia2022}. This focus is motivated by both their practical impact and the limited prior work on this type of noise: to the best of our knowledge, only two studies have explicitly examined this problem~\cite{8030620,Jia2022}.

We propose \emph{\app}\footnote{\label{fn:v1} A previous version of the tool was called LogCleaner and was introduced in an unpublished report \url{https://arxiv.org/abs/2004.07194v2} that followed a different evaluation plan.}, a downstream-task-agnostic technique for identifying and removing free-standing log messages. \app extends the heuristic originally proposed by~\citet{8030620} by accounting for dependencies between log messages. Specifically, \app computes dependency scores based on message co-occurrences and applies a clustering-based segmentation to automatically separate free-standing messages from regular ones. The underlying intuition is that randomly interleaved free-standing messages are likely to have lower dependency than regular messages.

As we have already implemented a prototype of \app, this registered report focuses on an evaluation plan to assess the impact of log cleaning with \app on two widely used downstream tasks: model inference (MI) and anomaly detection (AD). These tasks exploit different aspects of log data: MI aims to reconstruct models of expected behavior, whereas AD focuses on identifying deviations from normal execution patterns. The evaluation plan examines both the effectiveness and efficiency of downstream analysis tools when applied to cleaned versus original logs.

To summarize, the main contributions of this paper are:

\begin{itemize}
	\item We propose \app, an approach for identifying and removing free-standing log messages in black-box settings.
	\item We present a comprehensive empirical evaluation plan to assess the benefits of cleaning the log in the context of AD and MI.\item We plan to publicly release a full replication package including source code and the evaluation data. \end{itemize}

The remainder of the paper is organized as follows. Section~\ref{sec:preliminary} introduces background concepts on logs and downstream log analysis tasks. Section~\ref{sec:technique} presents \app. Section~\ref{sec:eval} describes the evaluation, with the analysis plan detailed in Section~\ref{sec:analysis_plan}. Threats to validity and related work are discussed in Sections~\ref{sec:threats-to-validity} and~\ref{sec:related-work}, respectively.  Section~\ref{sec:conclusion} concludes the paper.
 \section{Background}
\label{sec:preliminary}

\subsection{Logs}

A log is a sequence of log entries. A log entry contains a timestamp (recording 
the time at which the logged event occurred) and a log message (with run-time 
information). A log message can be further decomposed~\cite{messaoudi2018search} 
into a fixed part called the event template, characterizing the event type, and a variable part, which
contains tokens filled at run time with the values of the event parameters.
For example, the first log entry of the example log $l_\mathit{eg}$ shown in Figure~\ref{fig:example-log} 
is composed of the timestamp \texttt{20180526:10:00:01}, 
the template \texttt{\textbf{send}} $v_1$ \texttt{\textbf{via}} $v_2$, and the set of parameter values 
$\{v_1 = \texttt{MSG1}, v_2 = \texttt{CH1}\}$. More formally, let $L$ be the set 
of all logs, $T$ be the set of all event templates, and $P$ be the set of all 
mappings from event parameters to their concrete values. A log $l\in L$ is a 
sequence of log entries $\langle e_1, \dots, e_n \rangle$, with $e_i = (\tau_i, 
t_i, p_i)$, $\tau_i\in \mathbb{N}$, $t_i\in T$, and $p_i\in P$, for $i=1,\dots,n$.
Using this formalism, log $l_\mathit{eg}$ in Figure~\ref{fig:example-log} 
would correspond to  $l_\mathit{eg} = \langle e_1, e_2, \dots, e_{11} \rangle$ where $\tau_1=$ 
\texttt{20180526:10:00:01}, $t_1 =$ \texttt{\textbf{send}} $v_1$ \texttt{\textbf{via}} $v_2$, $p_1 = \{v_1 = 
\texttt{MSG1}, v_2 = \texttt{CH1}\}$, and so on. In the rest of the paper, we denote a template using 
its first word for simplicity; for example, we write the template 
\texttt{\textbf{send}} instead of the template \texttt{\textbf{send}} $v_1$ \texttt{\textbf{via}} $v_2$.

\begin{figure}[h]
	\small
	\centering
	\begin{tabular}{ll}
		\toprule
		$l_\mathit{eg}$ & Log entry (timestamp + message) \\
		\midrule
		$e_1$ & \texttt{20180526:10:00:01 \textbf{send} MSG1 \textbf{via} CH1} \\
		$e_2$ & \texttt{20180526:10:00:03 \textbf{check} MSG1} \\
		\rowcolor{lightgray} 
		$e_3$ & \texttt{20180526:10:00:05 \textbf{memory} OK} \\
		$e_4$ & \texttt{20180526:10:00:05 \textbf{send} MSG1 \textbf{via} CH1} \\
		$e_5$ & \texttt{20180526:10:00:06 \textbf{check} MSG1} \\
		$e_6$ & \texttt{20180526:10:00:07 \textbf{send} MSG1 \textbf{via} CH1} \\
		$e_7$ & \texttt{20180526:10:00:08 \textbf{check} MSG1} \\
		$e_8$ & \texttt{20180526:10:00:09 \textbf{send} MSG1 \textbf{via} CH1} \\
		\rowcolor{lightgray} 
		$e_9$ & \texttt{20180526:10:00:10 \textbf{memory} OK} \\
		$e_{10}$ & \texttt{20180526:10:00:10 \textbf{check} MSG1} \\
		\bottomrule
	\end{tabular}
	\caption{Logs of the running example (free-standing log messages highlighted in gray)}
	\label{fig:example-log}
\end{figure}

The example log $l_\mathit{eg}$ shown in Figure~\ref{fig:example-log} will 
be used as a running example throughout the paper. It has three templates, 
i.e., $T=\{\texttt{\textbf{send}}, \texttt{\textbf{check}}, 
\texttt{\textbf{memory}}\}$. Among them, 
\texttt{\textbf{send}} and \texttt{\textbf{check}} are regular event
templates that represent the functional behavior of the system, while 
\texttt{\textbf{memory}} is a free-standing event template 
that represents the operational state of the system. Free-standing messages 
are highlighted in gray; as visible from the figure, they are randomly interleaved with regular messages.

In practice, a log file is often a sequence of free-form text lines rather than a sequence of structured log entries. However, automatic log parsing has been widely studied to decompose free-form text lines into structured log entries by accurately identifying fixed parts (i.e., log message templates)~\cite{Zhang2023_survey}. For this reason, throughout the paper, we assume that logs are given in a structured form.

\subsection{Downstream log analysis tasks}
\label{sec:downstream_tasks}

In this paper, we assess the effect of log cleaning on two downstream tasks: anomaly detection and model inference.

\subsubsection{Model Inference} 

Model inference aims to extract behavioral models---typically in the form of Finite State Machines (FSMs)---from the execution logs of software systems. These models support a variety of software engineering tasks, including program comprehension~\cite{Cook:1998:287001}, test case generation~\cite{6200086}, and model checking~\cite{clarke2018model}. The task is particularly relevant because accurate software models are often missing or become outdated as systems evolve, while execution logs provide an accessible source of observed system behavior.

However, inferring accurate models presents several challenges. Execution data may vary greatly in quality and volume, logs may contain noise or incomplete traces, and finding the right balance between generality and precision is non-trivial. These issues make automated model inference both necessary and difficult, especially for large and complex systems.

A wide range of techniques have been proposed to address these challenges, ranging from classical algorithms for inferring FSMs~\cite{biermann1972synthesis,Beschastnikh2011Lev,LUO201713} to richer variants such as guarded FSMs (gFSMs)~\cite{walkinshaw2016inferring,mariani2017gk} and probabilistic extensions~\cite{Emam2018Inf}. More recent approaches include component-based inference techniques~\cite{Shin2022} and large language models~\cite{Wei2025}.

\subsubsection{Log-based Anomaly Detection}

Log-based anomaly detection uses execution logs to identify deviations from normal behavior that may signal faults, security breaches, or other system issues~\cite{he2021survey}. A typical anomaly detection pipeline consists of three steps: \begin{inparaenum}[(1)]
	\item log parsing to extract structured events from raw log lines,
	\item log partitioning to group events into sequences, and
    \item anomaly detection to identify anomalous sequences based on their event patterns.
\end{inparaenum}

Log-based anomaly detection spans approaches from statistical methods to deep learning~\cite{Landauer2023,Ali2025,Chandola2009}. The appropriate choice is context-dependent: deep learning models tend to perform best with abundant, complex data, while traditional machine learning models can be more suitable under data scarcity or when interpretability is required.
 \section{Approach}
\label{sec:technique}

The goal of \app is to identify free-standing event templates in a set of logs and to remove from the logs the messages corresponding to these
templates. For our running example shown in
Figure~\ref{fig:example-log}, this means that \app should identify the
template \texttt{\textbf{memory}} as
free-standing and remove the corresponding messages, highlighted in gray.

The intuition behind \app is that free-standing (event) templates are
distinguishable from regular (event) templates by looking at the
\emph{dependency} of the corresponding messages
in the logs. For instance, in our running example, the \texttt{\textbf{memory}}
template is distinguishable from the other templates
(\texttt{\textbf{send}} and \texttt{\textbf{check}}), because
\texttt{\textbf{send}} is frequently followed by \texttt{\textbf{check}}
and \texttt{\textbf{check}} is frequently preceded by
\texttt{\textbf{send}}. However,  template \texttt{\textbf{memory}} is neither frequently followed nor preceded by the other templates.
Assuming there is a way to measure, in the logs recorded during the execution of the system, the degree of dependency between templates, we expect
free-standing templates to have a much lower dependency score on other templates than regular templates.

Based on these observations,  \app  includes two main
steps: \emph{dependency score calculation} and \emph{clustering-based segmentation}.
The former aims at computing
the degree of dependency among templates based on their co-occurrences.
The latter automatically
partitions free-standing templates and regular templates based
on the dependency score of each template on the others.

Algorithm~\ref{alg:dependency} shows the pseudo-code of \app.
It takes as input a set of logs $L$ and a set of templates $T$; it
returns a set of cleaned logs $L_\mathit{cl}$, in which the free-standing messages
(i.e., the messages having free-standing event templates) have been removed.

\begin{algorithm}
	\SetKwInOut{Input}{Input}
	\SetKwInOut{Output}{Output}
	
	\Input{Set of Logs $L$ \\
		Set of Templates $T$}
	\Output{Set of Logs $L_\mathit{cl}$}
	
	Map from $T$ to Set of Reals $\mathit{mScore}$ \label{alg:dep:init}\\
	\ForEach{$x\in T$}{\label{alg:dep:beginX}
		$\mathit{mScore}[x] \gets 0$ \\
		\ForEach{$y\in T \setminus \{x\}$}{\label{alg:dep:beginY}
			$\mathit{mScore}[x] \gets \max(\mathit{mScore}[x], \mathit{dScoreCalc}(x, y, L))$\label{alg:dep:dScore}\\
		}\label{alg:dep:endY}
	}\label{alg:dep:endX}
	Set of Templates $T_\mathit{fs} \gets \mathit{clusterBasedSegment}(T, mScore)$\label{alg:dep:seg}\\
	Set of Logs $L_\mathit{cl} \gets \mathit{removeMessagesOf}(T_\mathit{fs}, L)$\label{alg:dep:rm}\\
	\textbf{return} $L_\mathit{cl}$\\
	
	\caption{\app}
	\label{alg:dependency}
\end{algorithm}

For each template $x\in T$, the algorithm determines the maximum value
of the dependency score (the value $\mathit{mScore}[x]$ for the key
$x$ in the associative array $\mathit{mScore}$), by computing the
individual dependency scores of $x$ on the other templates
$y \in T \setminus \{x\}$ in $L$
(lines~\ref{alg:dep:beginX}--\ref{alg:dep:endX}). This last step is
done by the \textit{dScoreCalc} function, described in detail in
subsection~\ref{sec:dScore}. We calculate the maximum value because, when $x$ is not 
free-standing, it has a high dependency score with at least one of the templates 
in $T-\{x\}$. Using the calculated $\mathit{mScore}$, the
algorithm calls the \textit{clusterBasedSegment} function (described
in detail in subsection~\ref{sec:clustering}) to identify the set of
free-standing templates $T_\mathit{fs}$ from $T$
(line~\ref{alg:dep:seg}). The algorithm ends by returning the set of
cleaned logs $L_\mathit{cl}$, obtained (line~\ref{alg:dep:rm}) by
removing the free-standing messages from $L$ based on the free-standing
templates in $T_\mathit{fs}$.

In the following subsections, we illustrate the two main steps:
dependency score calculation and clustering-based segmentation.

\subsection{Dependency Score}
\label{sec:dScore}

To measure the dependency score of a template $x$ on another template
$y$ for a set of logs $L$, we consider not only the dependency for
which $x$ could be a \emph{cause} of $y$ in $L$ but also the
dependency for which $x$ could be a \emph{consequence} of $y$ in $L$.
More precisely, we define the \emph{forward dependency score} of $x$
on $y$ for $L$, denoted with $\mathit{dScore}_f(x, y, L)$, as a
measure of how likely an occurrence of $x$ is \emph{followed} by an
occurrence of $y$ (i.e., $x$ is a cause of $y$) throughout $L$.
Similarly, the \emph{backward dependency score} of $x$ on $y$ for $L$,
denoted with $\mathit{dScore}_b(x, y, L)$, is a measure of how likely
an occurrence of $x$ is \emph{preceded} by an occurrence of $y$ (i.e.,
$x$ is a \emph{consequence} of $y$) throughout $L$. Since
$\mathit{dScore}_b(x, y, L)$ is equivalent to
$\mathit{dScore}_f(x, y, \mathit{rev}(L))$, where $\mathit{rev}(L)$ is the set of reversed logs of $L$. Below, we present only the algorithm to compute the forward dependency score.

First, we introduce the concept of a log entry occurring
after another one. More formally, given a log entry $e_x$ of a
template $x$ in a log $l$, we say that a log entry $e_y$ of a template
$y$ is the \emph{first-following} log entry for $e_x$ in $l$ if $e_y$
is the first log entry of $y$ between $e_x$ and the next log entry of
$x$ in $l$. For instance, in our
running example log $l_\mathit{eg}$, the log entries of the
\texttt{\textbf{memory}} template are $e_3$ and $e_9$
while the log entries of the \texttt{\textbf{send}}
template are $e_1$, $e_4$, $e_6$, and $e_8$. The log entry $e_4$
(of template \texttt{\textbf{send}}) is the first following log entry
for $e_3$ (of template \texttt{\textbf{memory}}),
because $e_4$ is the first log entry of \texttt{\textbf{send}}
between $e_3$ and $e_9$. However, there is no 
first-following log entry for $e_9$ because there is no log entry of
\texttt{\textbf{send}} between $e_{10}$ and the end of the log.

However, simply checking whether there is a first-following log entry
is not enough to compute the dependency score, because it does not
consider \emph{how close} the two log entries are. To take into account the distance between log entries, we define the
\emph{co-occurrence score} between two log entries $e_x$ and $e_y$ in
a log $l$, denoted with $\mathit{cScore}(e_x, e_y, l)$, as
$$ \mathit{cScore}(e_x, e_y, l) = \frac{1}{\mathit{distance}(e_x, e_y,
	l)}$$
    where $\mathit{distance}(e_x, e_y, l)$ is the difference of
the indexes between $e_x$ and $e_y$ in $l$. For our running example
log $l_\mathit{eg}$, we have
$\mathit{cScore}(e_3, e_4, l_\mathit{eg}) = 1$ because
$\mathit{distance}(e_3, e_4, l_\mathit{eg})=1$.
When an entry $e_x$ has no first-following log entry,
as it is the case for $e_9$ in the above example,
we have $\mathit{cScore}(e_x, \mathit{NaE}, l) = 0$ where $\mathit{NaE}$ indicates ``Not an Entry''.

We can then compute the dependency score between two templates $x$ and $y$
for a set of logs $L$ as the average of the
$\mathit{cScore}$ values of all log entries of $x$ with its first-following
log entries of $y$. More formally, we have 
$$
	\mathit{dScore}_f(x, y, L) =
	\frac{\sum_{l \in L}{\sum_{e_x \in E_{x,l}}{\mathit{cScore}(e_x, e_y, l)}}}{n}
$$
where $E_{x,l}$ is the set of log entries of $x$ in $l$,
$e_y$ is the first-following entry of $y$ for $e_x$ in $l$,
and $n$ is the total number of log entries of $x$ in $L$.
In this way, we measure how likely and how closely an occurrence of $x$ is
followed by an occurrence of $y$ throughout $L$.
A value of $\mathit{dScore}_f(x, y, L) = 1$ indicates that $x$ always
immediately causes $y$ in the logs in $L$,
while a value $\mathit{dScore}_f(x, y, L) = 0$ indicates that $x$ cannot cause $y$ in the logs in $L$.
For the running example above, we have:
\begin{align*}
	\mathit{dScore}_f&(\texttt{\textbf{memory}}, \texttt{\textbf{send}},\{l_\mathit{eg}\}) \\
	&= \frac{\sum_{l \in \{l_\mathit{eg}\}}{\sum_{e_x \in E_{x,l}}{\mathit{cScore}(e_x, e_y, l)}}}{2} \\
	&= \frac{ \sum_{e_x \in \{e_3, e_9\}} {\mathit{cScore}(e_x, e_y, l_\mathit{eg})} }{2} \\
	&= \frac{1}{2} \times \left[ \mathit{cScore}(e_3, e_4, l_\mathit{eg}) + \mathit{cScore}(e_9, \mathit{NaE}, l_\mathit{eg}) \right] \\
	&= \frac{1}{2} \times \left[ 1 + 0 \right] = 0.5
\end{align*}

We recall that we compute both the forward and backward dependency scores. The function $\mathit{dScoreCalc}(x, y, L)$ in Algorithm~\ref{alg:dependency} returns the maximum between $\mathit{dScore}_f(x, y, L)$ and $\mathit{dScore}_b(x, y, L)$. As a result, for our running example, we have:
\begin{align*}
	\mathit{mScore}[\texttt{\textbf{memory}}] = \max (&\mathit{dScore}_f(\texttt{\textbf{memory}}, \texttt{\textbf{send}},\{l_\mathit{eg}\}), \\
	&\mathit{dScore}_b(\texttt{\textbf{memory}}, \texttt{\textbf{send}},\{l_\mathit{eg}\}), \\
	&\mathit{dScore}_f(\texttt{\textbf{memory}}, \texttt{\textbf{check}},\{l_\mathit{eg}\}), \\
	&\mathit{dScore}_b(\texttt{\textbf{memory}}, \texttt{\textbf{check}},\{l_\mathit{eg}\}) ) \\
	= \max(&0.5, 0.75, 0.75, 0.75) = 0.75
\end{align*}
In the example, $\mathit{mScore}$ for the free-standing template \texttt{\textbf{memory}} 
is relatively high (0.75): this happens simply because of the small
size of our running example (where the total number of templates is three). For the regular templates \texttt{\textbf{send}} 
and \texttt{\textbf{check}}, we have $\mathit{mScore}[\texttt{\textbf{send}}] = 0.875$ 
and $\mathit{mScore}[\texttt{\textbf{check}}] = 0.875$.
The following subsection describes how to automatically distinguish free-standing 
templates and regular templates based on the $\mathit{mScore}$ values.

\subsection{Clustering}
\label{sec:clustering}
As mentioned earlier, the (maximum) dependency score
$\mathit{mScore}$ value of free-standing templates is likely to be less than
that of regular templates. Furthermore, in our preliminary
experiments, we observed that the gap among the $\mathit{mScore}$
values for free-standing templates is often smaller than the gap
between the highest $\mathit{mScore}$ value of a free-standing
template and the lowest $\mathit{mScore}$ value of a regular
template. This suggests that the set of free-standing templates could
form a cluster based on $\mathit{mScore}$. Therefore, we propose a
heuristic to partition free-standing templates and regular
templates using clustering.

We first generate multiple clusters of templates based on the value
$\mathit{mScore}$. The number of clusters can be more than two because
regular templates often lead to multiple clusters. Since the
number of clusters is not known in advance, we use the Mean-Shift
clustering algorithm~\cite{comaniciu02:_mean}.
Among the generated clusters, the cluster with the smallest
$\mathit{mScore}$ value is assumed to be one of the free-standing templates.

For instance, if we apply the clustering-based segmentation heuristic
to our running example, with
$T = \{ \texttt{\textbf{send}}, \texttt{\textbf{check}},
\texttt{\textbf{memory}} \}$ and using the $\mathit{mScore}$ values
computed in subsection~\ref{sec:dScore}, the clustering algorithm will
generate two clusters $c_1 = \{ \texttt{\textbf{send}}, \texttt{\textbf{check}} \}$,
and $c_2 = \{ \texttt{\textbf{memory}} \}$. Since $c_2$ has the smallest
$\mathit{mScore} = 0.75$, the \texttt{\textbf{memory}} template in
$c_2$ is identified as free-standing.

After identifying free-standing templates with the clustering-based
segmentation heuristic, the dependency analysis algorithm ends with
the removal of the log entries containing one of the identified
free-standing templates. In the case of our running example log
$l_\mathit{eg}$, this means removing the entries with template \texttt{\textbf{memory}}; the final, cleaned version of the log is the one
without any of the free-standing messages highlighted in grey
in Figure~\ref{fig:example-log}.

 \section{Evaluation Design}
\label{sec:eval}

In this section, we present the plan for evaluating the impact, on the effectiveness and efficiency of downstream tools, of cleaning free‑standing templates from log data using \app. We focus on two common but distinct log analysis tasks: MI and AD (see Section~\ref{sec:downstream_tasks}). Both tasks rely heavily on the structure and consistency of logs, but are impacted by noise in different ways.

The evaluation is guided by the following research questions:
    
\begin{description}
    \item[RQ1.] \textit{How does cleaning the log with \app affect the effectiveness of downstream task tools?}
    \item[RQ2.] \textit{How does cleaning the log with \app affect the efficiency of downstream task tools?}
\end{description}

These research questions capture two key ways in which free‑standing messages can influence downstream analyses. RQ1 focuses on effectiveness, as noise and randomly interleaved events can reduce the accuracy of AD and MI techniques, while RQ2 focuses on efficiency, since large log volumes increase computational cost. Balancing these two dimensions is critical in practice, given that real‑world logs are typically large and noisy. As downstream tasks exploit log data differently, we do not expect cleaning to have a uniform effect across tasks.

\subsection{General Evaluation Procedure}
\label{sec:general_eval_procedure}

We follow the same evaluation pipeline for both RQs: we prepare $L_{\mathit{org}}$, produce $L_{\mathit{cl}}$ using \app, run the downstream tool on both logs, and compare outcomes using the planned descriptive and statistical analysis. All experiments will be conducted on a high‑performance computing platform.
 
For baseline comparison, we evaluate against LogSed~\cite{8030620} and LogBoost~\cite{Li2026}. Although other log cleaning approaches have been proposed (see Section~\ref{sec:related-work}), their implementation is unavailable~\cite{conforti2017, zhang2024, Locke2022}. We also investigate their false positive rates, i.e., how often relevant messages are removed incorrectly.

\subsection{Experimental Setup - Model Inference}
\label{sec:eval_MI}

\noindent\textbf{Tool.} 
Our evaluation will use only one tool for MI, as we do not aim to benchmark MI approaches; specifically, we chose MINT~\cite{minttool} as it is state-of-the-art, publicly available, and has been shown to be accurate~\cite{Shin2022}.

\noindent\textbf{Datasets and Data Preparation.}
We plan to assemble a benchmark of logs obtained from the execution of at least 11 publicly available system models (in the form of Finite State Machines - FSM). These systems have been proposed previously and widely used in the MI literature~\cite{4023976,5609576,LO20122063}. We will use the methodology proposed by~\citet{nimrod} to generate logs from these FSM models, using the publicly available trace generator by~\citet{LO20122063}, configured to provide state coverage of four visits per state and a minimum of 1000 logs. This configuration is expected to provide a sufficiently large dataset of execution logs to support the planned statistical analysis.

Logs generated from FSM models are, by construction, purely regular. To study the effect of cleaning noisy logs, we introduce noise by randomly interleaving free‑standing messages with regular messages. Specifically, we generate log entries corresponding to $n$ free‑standing event templates and inject them into the logs by randomizing their timestamps. The value of $n$ is chosen not to exceed the number of regular templates in the generated logs. To assess the performance of \app under varying noise conditions, we vary the noise rate (NR)---defined as the proportion of injected free‑standing log entries over the total number of log entries---between 0.1 and 0.9.

\noindent\textbf{Evaluation process.}
As MI tools infer system behavior from logs, their accuracy depends on how well the learned model matches the reference model. We assess this using the deterministic language-cardinality method of~\citet{clun2024rigorous}, which enumerates all traces up to a user-defined maximum length $k$, chosen by domain knowledge or set sufficiently large to cover relevant behaviors.

Additionally, to account for randomness in the generation of $L_\mathit{org}$ and the injection of free-standing messages, we repeat the model assessment 30 times for each system in our benchmark.

\noindent\textbf{Collected Metrics.}
We measure \textit{effectiveness} using cumulative precision and recall of inferred models~\cite{clun2024rigorous}.

The \textit{efficiency} is measured as the execution time of MINT in seconds.

\subsection{Experimental Setup - Anomaly Detection}
\label{sec:eval_AD}

\noindent\textbf{Tools.}
We will use invariant mining (IM)~\cite{Lou2010} and one-class SVM (OC-SVM)~\cite{Scholkopf2001}. IM is an unsupervised method, whereas OC-SVM is semi-supervised; both models are trained using only normal execution logs. This choice reflects practical settings in which fully labeled logs are scarce. We exclude deep learning‑based methods due to their high computational cost and lack of consistent performance gains over simpler machine learning techniques~\cite{Ali2025}. To reduce implementation bias, we will adopt Loglizer~\cite{he_experience_2016}, a publicly available and widely used toolkit for log-based AD.

\noindent\textbf{Datasets and Data Preparation.}
Following recent studies~\cite{LeZhang2021NeuralLog}, we will include at least the BGL, Thunderbird, and Spirit log datasets in our evaluation. These datasets are widely used in the AD literature and represent real-world logs collected from large-scale systems~\cite{Oliner2007,Landauer2024}. Additional datasets may be included if suitable data are identified; nevertheless, these three datasets provide a reasonable and well-established basis to evaluate the impact of \app.

As in recent work~\cite{he_experience_2016,Ali2025}, we will use Drain~\cite{8029742} to extract templates and 
fixed time windows to segment logs into sequences for AD. To comprehensively evaluate the impact of \app, we consider seven time window sizes (60, 100, 120, 300, 600, 1800, and 3600 seconds), as the choice of time window can affect the performance of AD methods~\cite{LeZhang2021NeuralLog}.

\noindent\textbf{Evaluation process.}

We preprocess both AD models identically. We sample 80\% of the \emph{normal} sequences from $L_{\mathit{org}}$ for training; the test set contains the remaining 20\% of normal sequences plus all \emph{anomalous} sequences.
To construct $L_{\mathit{cl}}$, we run \app on the training data to identify free-standing templates, then remove all matching log messages from both the training and test sets.

\noindent\textbf{Collected metrics.}
To assess the \textit{effectiveness} of \app on AD, we measure precision, recall, and F1-score of the selected AD tools, following common practice~\cite{Ali2025,he_experience_2016}. 

For \textit{efficiency}, we measure the training time of the tools in seconds.

\section{Analysis Plan} 
\label{sec:analysis_plan}

\subsection{Analysis on the Effectiveness (RQ1)}

To assess the effect of log cleaning on the outcome measures of downstream tasks, we analyze AD and MI independently and assess each outcome measure collected on its own. In all analyses, the \textit{dependent variable} is the downstream task outcome measure, while the \textit{independent variable} indicates whether the input log is cleaned or not. 

As each system is evaluated using several configurations (e.g., different tools, noise rates, or window sizes) and each configuration is tested using the original and cleaned logs, the resulting data consist of repeated paired measurements that are grouped by system. 
Consequently, the primary analysis method will be linear mixed-effects models (LMMs)~\cite{LMM_book}. We account for system-level variability by including the system as a random effect, model log cleaning as the fixed effect of primary interest, and include factors relevant to each downstream task (e.g., noise rate and time window size) as additional fixed effects. The appropriate analysis technique depends on the observed characteristics of the data; therefore, we will apply transformations when appropriate and rely on generalized models if the assumptions of LMMs are not met. 

As the study is exploratory and we are primarily interested in the overall effect of cleaning, we summarize the results using marginal effects~\cite{wooldridge2010}. They describe how the expected value of an outcome changes when the explanatory variable of interest changes, while the other variables in the model remain constant. For MI, marginal effects aggregate results across different noise rates, while for AD they are aggregated across tools and time window configurations.

To complement the marginal effects, we will examine high-level trends associated with individual covariates, without analyzing each of them in depth.

\subsection{Analysis on the Efficiency (RQ2)}
\label{sec:RQ2}

We analyze execution/training time differences using the same analysis procedure as described for RQ1. In the applied models, all predictors and random effects remain unchanged; only the dependent variable differs. For RQ2, it corresponds to the execution/training time of the downstream analysis tool. 

The execution time of log-based analysis tools typically depends on the size of the input data. As log cleaning reduces the size of input logs, observed differences in execution or training time may be associated with this reduction. To assess this, we perform an exploratory mediation analysis~\cite{hayes2017introduction}. This analysis estimates the portion of the observed association that is statistically associated with the reduction in the number of templates in the input data.

 \section{Threats to validity}
\label{sec:threats-to-validity}

\textbf{External Validity.} Our evaluation focuses on two downstream tasks with distinct ways of exploiting log data. Consequently, our findings may not generalize to all types of log‑based downstream tasks. In particular, some tasks may benefit less from removing free‑standing messages or require different cleaning strategies. Nevertheless, MI and AD are common tasks, meaning that our findings remain relevant for practical log‑analysis scenarios.

\textbf{Internal Validity.} 
A primary threat comes from the datasets used in our evaluation. For MI, we rely on synthetically generated logs, as finding a reliable annotated real-world dataset of sufficient complexity is challenging. To make the data more realistic, we vary the noise rate and repeat the data generation 30 times for each system. 

For AD, we have identified three widely used public datasets. Although this enables comparison with prior work, the limited number of independent projects may reduce statistical power. As a result, our statistical analysis should be interpreted as exploratory; we focus on descriptive statistics and marginal effects rather than individual model coefficients.

Additional threats arise from log pre-processing. In the absence of session identifiers, we partition logs using fixed time windows, which may affect performance; we mitigate this by evaluating multiple window sizes. We also rely on a single standard log parser and keep templates unchanged; different parsers may produce templates with different granularity, potentially altering dependency scores and the resulting set of filtered templates.
Finally, results may depend on the chosen downstream tools, though we expect overall trends to generalize to comparable approaches.

Regarding baselines, LogSed and LogBoost were developed to improve AD rather than MI, so their results when applied to MI may reflect a task mismatch; we include them for consistency but interpret these findings cautiously. LogSed may still be relevant, as its template-removal heuristic could also benefit MI. Because the LogSed code is unavailable, we re-implement it based on the original paper to minimize implementation bias.

\textbf{Construct Validity.} 
We do not perform an empirical evaluation of \app itself. Conducting such an evaluation would require a reliable ground truth identifying which log messages are free‑standing; however, it is not possible to define such a ground truth unambiguously in a task‑agnostic manner. Ultimately, \app is not intended to perfectly clean logs; its goal is to improve downstream task performance, which is directly assessed in our study.

 \section{Related work}
\label{sec:related-work}

The closest work to \app is the automated free-standing message filtering technique by~\citet{8030620} (which we call LogSed). While its high-level intuition is similar to \app, LogSed models only one-way dependencies and requires user-set parameters, including thresholds that determine the size of the vicinity window used in its dependency analysis, whereas \app models two-way dependencies and is fully automatic.

In process mining, \citet{conforti2017} treat noise as weakly supported \emph{transitions}, i.e., infrequent directly-follows relations between templates. In contrast, \app targets weakly supported \emph{templates} by identifying templates that are structurally independent in the inferred dependency structure.

More recent AD-oriented methods include LogBoost~\cite{Li2026}, which removes templates shared by normal and anomalous sequences, and LogCleaner~\cite{zhang2024}, which prunes non-beneficial events (anti-events or duplicative events). LogAssist~\cite{Locke2022} instead targets redundancy for summarization by collapsing repeated events and compressing recurring, co-occurring subsequences. Overall, LogBoost and LogCleaner define noise relative to anomaly detection, while LogAssist defines it relative to summarization. In contrast, \app uses a task-agnostic notion of noise that is not tied to a single downstream task and filters independently occurring templates that are unlikely to reflect functional system behavior.

 \section{Conclusion}
\label{sec:conclusion}

In this registered report, we have presented \app, a task-agnostic log-cleaning tool that identifies and removes \emph{free-standing} messages from log files in a black-box setting by computing co-occurrence-based dependency scores for log message templates. The goal of \app is to enhance downstream task effectiveness and efficiency by reducing the noise in log files.  

We have presented an empirical evaluation plan of \app on two downstream tasks: model inference and anomaly detection. The plan specifies the downstream task tools, datasets, data pre-processing steps, evaluation metrics, and statistical analysis procedures.

\section*{Acknowledgements}
This research was funded in whole or in part by the Luxembourg National Research Fund (FNR), grant reference C22/IS/17373407/LOGODOR.

\footnotesize{
\bibliographystyle{IEEEtranN}

}

\end{document}